# Multi-resolution analysis of the H.E.S.S. Galactic Survey Sources and Search for Counterparts in CO and HI data.

A.Lemiere, R.Terrier, A.Djannati-Ataï,
for the H.E.S.S. Collaboration
*APC College de France, 11 place Marcelin Berthelot, Paris, France*
Presenter: A.Lemiere (alemiere@cdf.in2p3.fr), fra-lemiere-A-abs1-og22-oral

From May to July 2004, the central radian of the Galactic Plane was scanned by the H.E.S.S. (High Energy Stereoscopic System) telescopes at energies above 200 GeV. This survey was performed from -3° to +3° in latitude, for a total of 230 hours, revealing eight new VHE sources at a significance level greater than 6 $\sigma$ (standard deviations). We present a multi-resolution analysis of these sources based on a continuous wavelet transformation (CWT). Using CO and HI data, we investigate the possible associations of the potential counterparts proposed in [1], with sites of enhanced interstellar matter density.

## 1. Introduction

The H.E.S.S. atmospheric Cerenkov telescope system has been operating at full sensitivity over a year above an energy threshold of 100 GeV. Thanks to its stereoscopic design and its large field of view (5°), H.E.S.S. is well suited for the detection and mapping of extended VHE gamma-ray sources (e.g., SNR RXJ1713.7-3946 [2]). A survey of the inner part of the Galactic plane was made with H.E.S.S. during 2004, searching for emission from SNRs, pulsars and pulsar wind nebulae (PWN), micro-quasars and also possible new classes of sources. Eight new sources have been discovered [1]. They cluster close to the Galactic plane and the majority of them show clear spatial extension.

Gamma-ray production through neutral pion decay may primarily take place at sites involving collision of SNR shock waves or OB type stellar winds with clouds or swept up interstellar matter. To investigate the possible associations of the potential point-like or extended counterparts at other wavelengths with sites of enhanced interstellar matter density, we have made a detailed study of the molecular and atomic matter distribution in the line of sight to each source, using CO and HI data.

## 2. H.E.S.S. first galactic scan results

The Galactic Plane scan was carried out for a total of 230 hours, in the galactic longitude and latitude ranges of 330° to 30° and -3° to +3°, respectively, reaching an peak flux sensitivity of about 2% of the Crab nebula above 200 GeV. The new sources have a significance level greater than $6\sigma$, with fluxes ranging from a few percent up to 25 percent of that of the Crab nebula. The narrow latitude distribution of the sources, with an average value of -0.25°, is similar to that of SNRs and pulsars, but as the parent population luminosity distribution is unknown, one can not make quantitative statements on their compatibility. Six of the new sources are significantly spatially extended, the two others showing a marginal extension with respect to the PSF (Point Spread Function) of the telescope system.

The excess maps of theses sources are shown in [3]. Alternative analysis techniques are being explored such as a multi-scale decomposition of the data with which the images in Figure 1 were obtained. The method consists of filtering the signal map with an isotropic wavelet at several given scales. In practise the convolution is made with a two-dimensional continuous wavelet [4]. At each scale, a given pixel is selected for the final image reconstruction if its pre-trial significance (evaluated using the distribution of the CWT of the background image) is greater than $S_h = 6.5\sigma$, or $S_l = 4.5\sigma$ if it is has a neighbouring pixel at a distance closer than the considered scale with $S > S_h$.

## 3. Search for atomic and molecular clouds in the ISM

The fact that all of the new H.E.S.S. sources cluster on the Galactic plane, and are rather concentrated in the central galactic arms' tangent region does suggest a possible relation with the massive star forming/high density ISM regions. The CO data from the composite survey [5] have been used for a systematic search for molecular clouds in the line of sight to the H.E.S.S. new sources. The (l,b,v) cube gives the carbon monoxide brightness temperature map as a function of latitude, longitude and velocity.

For each source, we integrate the total CO emission over the source extension in (l,b). Peaks in the resulting velocity profiles are then selected by requiring an intensity higher than $1.5\text{K.deg}^2$. The extension of the velocity-selected region $\delta V$ is defined as the FWHM of a Gaussian fit to the peak ($\delta V$ represents also the internal motion of the cloud). For each peak, a map is produced by integrating the (l,b,v) cube on $\delta V$ for a field of view of $\pm 3°$ around the source. The H.E.S.S. excess contours are superimposed on these maps and potential spatial correlations are then searched visually.

The cloud selection criteria include also the spatial coincidence with potential counterparts and compatibility of their distance (when known) with one of the distance solutions for the cloud. These criteria are not sufficient to claim any firm association, but are intended as an aid to counterpart identification. A simple test has been made to evaluate the chance probability of spurious spatial coincidences by reversing the longitude of the H.E.S.S map. It yields a probability of $\approx 0.625$ for chance associations.

An empirical Galactic rotation curve model by [6] is used to estimate the distances kinematically. The rotation curve is computed for each source's galactic longitude, and when associated to each velocity-peak, yields two solutions. Cloud masses are determined with CO luminosities, on the empirically- based assumption that the integrated CO line intensity is proportional to the column density of $H_2$ along the line of sight and on the additional assumption that the cloud's hydrogen is entirely molecular. Ref [7] has been used to evaluate the molecular column density, the mass and the effective extension of the molecular clouds (noted $R_{\text{eff}}$). The HI (l,b,v) cube data from the Southern Galactic Plane Survey at 21cm has also been used with the same method as above, but as atomic matter is more diffuse than CO, the identification of structures is more difficult. In practise, we have searched in these data for the velocity ranges which correspond to a relevant molecular structure found in the CO survey.

## 4. Discussion of H.E.S.S. source counterparts

HESS J1614-518 is an extended and elongated source close to HESS J1616-508. It is one of the brightest sources detected, but no counterpart in this region at any wavelength has been found.
HESS J1616-508 is an extended source with a rather circular shape and a relatively high flux. HESS J1616-508 is close to the SNRs G 332.4-0.4 (RCW103) and G 332.4-0.1(Kes32) but not fully coincident. A compelling possible counterpart is the young and energetic hard X-ray pulsar PSR J1617-5055 [1] which has an estimated distance of 6.5 kpc. The detailed CO study described here has revealed four dense molecular regions in the line of sight. One of them ($\delta V = [-55, -46]$ km/s, $R_{\text{eff}} \approx 20$ pc) shows an interesting spatial coincidence with the H.E.S.S. source. The shorter-distance solution is compatible with the pulsar distance (5.5 kpc), with a total mass $M \approx 5.1 \times 10^5 M_{\text{sol}}$. This region corresponds also to a high HI density, but no characteristic structure has been seen.

HESS J1640-465 is marginally extended with a quasi-symmetrical shape. The most promising counterpart is the broken shell SNR (G 338.3-0.0), also detected by ASCA (AX J1640-4632) with a flux of $1.2 \times 10^{-12}$erg cm$^{-2}$s$^{-1}$ and a soft X-ray spectrum (index of 2.8) in the energy range $(0.7-7)$ keV, typical of SNR hot plasma emission with T $\approx 0.2-2$ keV [8]. This SNR is on the edge of a large HII region located in the Norma arm tangent. Nine dense molecular regions have been found in the line of sight, one having an interesting spatial coincidence with the H.E.S.S. source ($\delta V = [-109, -95]$ km/s, R$_{eff} \approx 22.3$ pc), a short-distance solution of (d $\approx 6.4$ kpc) and a total mass of M $\approx 7 \times 10^5$M$_{sol}$.

HESS J1804-216 is an extended source. It is spatially coincident with the south-western part of SNR G8.7-0.1(W30) detected during the ASCA survey [8]. A young Vela-like pulsar PSR J1803-2137 with a spin-down age of 16 k-years is coincident with the edge of the intense region. Although [9] report radio continuum evidence for massive star formation and ultra-compact HII regions around W30, we did not find any compelling association with the H.E.S.S. source.

HESS J1813-178 is one of the less extended sources. This source has a compelling positional coincidence with an unidentified ASCA source AGPS273.4-17.8 [8]. It is also at a distance of 10' from the radio source W33 which is an ultra-compact HII region with traces of recent star formation [1]. Seven molecular cloud have been found in the line of sight, one of them being in spatial coincidence with the source ($\delta V = [96, 115]$ km/s, R$_{eff} \approx 17.2$ pc) and a high velocity dispersion of $\sigma \approx 13.9$ km/s. Its close distance solution is estimated at d $\approx 6.6$ kpc and its total mass is around $M \approx 3.4 \times 10^5 M_{sol}$.

HESS J1825-137 is a quite extended source. We can see in figure 1(d) an unidentified EGRET source 3EG J1825-1302 [1] at 43 arc minutes from these source, but with low probability of association. A young-Vela like pulsar PSR B1823-13 (PSR J1826-1334) at a distance d = 3.9 kpc (seen in X-ray with XMM) has an interesting position on the edge of the source, near the maximum of emission. The CO study has shown seven high-density molecular regions in the line of sight of the source. One of them ($\delta V = [45, 53]$ km/s, R$_{eff} \approx 14.3$ pc) has a compelling spatial correlation with the pulsar and the intense region of the H.E.S.S. source. The close-distance solution (4 kpc) gives a total mass around M $\approx 3.9 \times 10^5$M$_{sol}$. It is in agreement with the pulsar distance. The large velocity dispersion ($\sigma = 10.3$km/s) could indicate a possible association of the source with the pulsar wind. The molecular cloud found here provides naturally the ISM inhomogeneity toward the north-south direction which is necessary to explain the asymmetry of the X-ray nebula. HESS J1834-087 is an extended source as well. J1834-087 has a compelling positional agreement with the SNR G23.3-0.3 (W41) at an estimated distance of 4.8 kpc. The 20 cm radio map contours show clearly the SNR broken shell with a hot spot in the centre. The H.E.S.S. excess is concentrated within the shell structure, as opposed e.g. to RX J1713-3946. A PWN-type association seems then more likely. A diffuse high intensity CO region is spatially coincident with the source. It has a compatible distance with the SNR but reveals no peculiar structure.

HESS J1837-069 is an extended source. Its most promising counterpart is an unidentified ASCA source (AXJ1838.0-0655)[8] (also detected by INTEGRAL and BeppoSAX). Eight molecular clouds have been found in the line of sight, one of them with a centroid compatible with that of the VHE Source ($\delta V = [0, 10]$ km/s, R$_{eff} \approx 4.3$ pc). For the close solution distance (d $\approx 0.5$kpc), the total mass is M $\approx 1.2 \times 10^6$M$_{sol}$.

## 5. Summary

The multi-wavelength search for counterparts for H.E.S.S. galactic survey sources has shown four plausible associations with shell or broken-shell SNRs, three pulsar wind nebula and two unidentified EGRET sources as possible counterparts [1] [3]. Images of the VHE regions, through an alternative multi-scale analysis of the data, have been obtained here. The systematic search for molecular clouds along the line of sight of the these sources shows some possible interactions with shock-waves either from SNR shells -in two or three cases- or from a pulsar-powered nebula in two cases. One of the sources, HESS J1614-518, has no obvious counterpart and may represent a new class of VHE emitter.

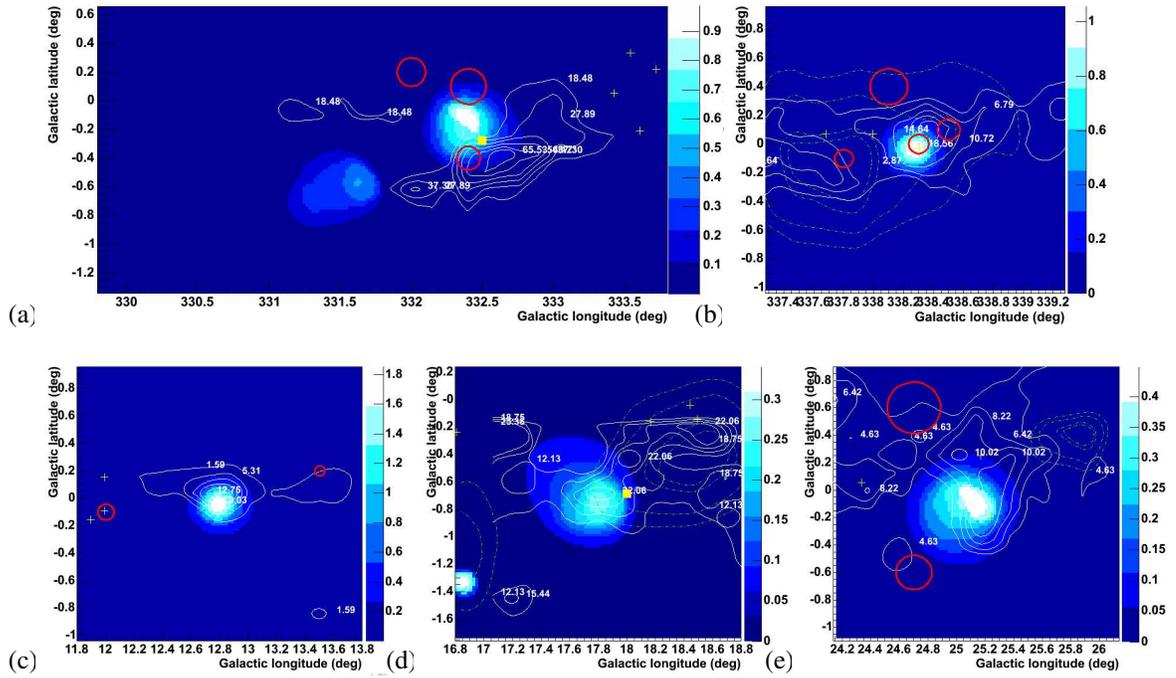

**Figure 1.** Multi-resolution filtered and reconstructed images of H.E.S.S.survey sources (the colour scale is arbitrary): (a) HESSJ1614-518 and HESS J1616-508 (b) HESSJ1640-465 (c) HESSJ1813-178 (d) HESSJ1825-137 (e) HESSJ1837-069. Possible counterpart are also indicated. The positional confidence contours for EGRET sources are shown in dash Gray (green), ASCA sources are indicated with crosses. The SNRs are indicated as circles (red) and the molecular cloud contours are shown in white, the numbers show column intensity levels in K.kms$^{-1}$ units; the pulsars are shown as light Gray squares.